\documentclass[smallcondensed]{svjour3}
\smartqed
\usepackage{graphicx}

\begin{document}

\title{Angular momentum exchange during secular migration of two\,-\,planet systems}

\titlerunning{Secular migration of two\,-\,planet systems} 

\author{Adri\'an Rodr\'iguez \and Tatiana A. Michtchenko \and Octavio Miloni}

\authorrunning{A. Rodr\'iguez et al.} 

\institute{A. Rodr\'iguez \and T. A. Michtchenko \at
Instituto de Astronomia, Geof\'{\i}sica e Ci\^encias Atmosf\'ericas, IAG\,-\,USP,\\
Universidade de S\~ao Paulo, S\~ao Paulo, SP, Brazil\\
\email{adrian@astro.iag.usp.br} \and  O. Miloni \at Facultad de Ciencias Astron\'omicas y Geof\'isicas, Universidad Nacional de La Plata, La Plata, Argentina\\
               }

\maketitle

\begin{abstract}

We investigate the secular dynamics of two-planet coplanar systems evolving under mutual gravitational interactions and dissipative forces. We consider two mechanisms responsible for the planetary migration: star-planet (or planet-satellite) tidal interactions and interactions of a planet with a gaseous disc. We show that each migration mechanism is characterized by a specific law of orbital angular momentum exchange. Calculating stationary solutions of the conservative secular problem and taking into account the orbital angular momentum leakage, we trace the evolutionary routes followed by the planet pairs during the migration process. This procedure allows us to recover the dynamical history of two-planet systems and constrain parameters of the involved physical processes.


\end{abstract}

\section{Introduction}\label{sec-0}

Recently, Michtchenko and Rodr\'iguez (2011) have purposed a new method for a
qualitative study of the planet migration originated by a generic dissipative mechanism.
It has been shown that, under assumption that the dissipation processes are sufficiently
slow, the evolutionary routes of migrating planets in the phase space are
traced by stationary solutions of the conservative secular problem. Therefore,
the modeling of planet migration consists in the calculation of families
composed by Mode I and Mode II stationary solutions, parameterized by the mass
ratio, for all possible values of planetary semi-major axes and orbital
angular momentum of the system.

It is presently accepted that the main mechanisms responsible for planet
migration are the star-planet (and planet-satellite) tidal interactions and
gravitational interactions of a planet with a gaseous disc.
Dissipative interactions remove (or add) orbital energy from the planetary system,
producing changes of semi-major axes. In the
case of tidal interactions, orbital energy is transformed into thermal energy,
which is dissipated in the interior of the tidally deformed body. Migration of
the system can be either toward or outward the central body, depending on the
properties of the tidal interactions and some parameters such as Love numbers
and quality factors (Darwin 1880; Jeffreys 1961; Goldreich and Soter 1966; Hut
1981; Dobbs-Dixon et al. 2004; Ferraz-Mello et al. 2008). During disc-planet
interactions, orbital energy is exchanged with a surrounding
gaseous protoplanetary disc and, generally, the planetary orbital decay occurs
(for a review, see Armitage 2010 and references therein). In addition,
migration can also occur through gravitational scattering between planets and a
remnant planetesimal debris (e.g. Fern\'andez and Ip 1984;
Kirsh 2009).

Dissipative forces also alter the orbital angular momentum of the
system. Indeed, tidal torques modify the rotational angular momenta of
the interacting bodies, which are transferred to the orbits of planets
or satellites (e.g. Mignard 1979; Correia et al. 2008). During
disc-planet interactions, the angular momentum is exchanged through
gravitational torques between the planets and the disc (e.g. Lin et al. 1996;
Goldreich and Sari 2003; Masset et al. 2006). In Michtchenko and Rodr\'iguez
(2011), we have purposed a generic approach to describe the exchange of the
orbital angular momentum of the system during migration. We have introduced the
orbital angular momentum leakage, defined as a portion $\alpha$ of the orbital angular momentum
variation produced by migration (i.e. expansion or shrinkage of the
planetary orbits), which is extracted (or added to) from the planet system.

In this paper, we show that each physical process is characterized by a
specific law of orbital angular momentum leakage. We show that $\alpha$
is a function of the eccentricity of the planet affected by
the dissipative force. Moreover, we show that $\alpha$ also depends on a set
of physical parameters related to planets, star and disc. Finally, owning the
characteristic $\alpha$-function, we construct the evolutionary routes of
the system for each specific mechanism driving the planet migration. This
approach provides us with a general idea of how the system could evolve under
a variety of migration conditions and physical models.

In Section 2 we briefly introduce some basic aspects of the secular dynamics
of two-planet systems evolving under dissipative forces. Section 3 addresses
tidal interactions, where we include the cases of a close-in planet and a
satellite interacting with their parent star and planet, respectively.
Migration driven by disc-planet interaction is discussed in Section 4. The
dissipation is modeled through a drag Stokes-like force. Finally, Section 5 is
devoted to conclusions.

\section{Some aspects of the secular dynamics under dissipation}\label{sec-1}

\subsection{Orbital angular momentum exchange}\label{sec-1-1}
We consider a three-body system consisting of a central star with mass $m_0$ and two coplanar planets with masses $m_1$ and $m_2$. We assume that the system is far away from any mean-motion resonance. Hereafter, the indices $i = 1, 2$ stand for the inner  and  outer  planets, respectively.

In the astrocentric reference frame, the orbital angular momentum of the system is given, up to second order in masses, as
\begin{equation}\label{Lorb}
L_{\rm{orb}}=m'_1\sqrt{a_1(1-e_1^2)}+m'_2\sqrt{a_2(1-e_2^2)},
\end{equation}
where $m'_i\equiv m_i\sqrt{G(m_0+m_i)}$, $a_i$ and $e_i$ are planetary semi-major axes and eccentricities, and $G$ is the gravitational constant.

The migration of planets is originated by dissipative forces which affect the energy and the orbital angular momentum of the planets and results in variations of their semi-major axes and eccentricities. If dissipation is sufficiently slow (its rate is smaller than the proper frequency of the secular motion), the long-term variations of the orbital elements can be separated  into two components: one is produced by secular interactions with the companion and the other is due to external interactions. Hence, we can write

\begin{equation}\label{aidot}
\dot{a}_i=\dot{a}_i\,^{\rm{sec}}+\dot{a}_i\,^{\rm{ext}},
\end{equation}
\begin{equation}\label{eidot}
\dot{e}_i=\dot{e}_i\,^{\rm{sec}}+\dot{e}_i\,^{\rm{ext}},
\end{equation}
where the indices ``sec" and ``ext" stand for secular and external contributions of the total variation of each element.


The secular theory provides that $\dot{a}_i\,^{\rm{sec}}=0$ (to first order in masses) and, as a consequence, $\dot{a}_i=\dot{a}_i\,^{\rm{ext}}$. We stress that, since we are studying the secular evolution of the system, short-period variations of the orbital elements (of the order of orbital periods) are omitted in the analysis. Thus, Equations (\ref{aidot})-(\ref{eidot}) should be considered as the averaged equations describing the long-term variations of semi-major axes and eccentricities. In addition, the contribution of resonant terms is also neglected.


Assuming that the planetary masses are unaltered during migration, the time variations of $L_{\rm{orb}}$, in terms  of  the  time  variation  of  the  planetary semi-major axes, $\dot{a}_i$, and eccentricities, $\dot{e}_i$, is written as
\begin{equation}\label{Lorbdot}
\dot{L}_{\rm{orb}}=\frac{m'_1\sqrt{1-e_1^2}}{2\sqrt{a_1}}\dot{a}_1-\frac{m'_1\sqrt{a_1}e_1}{\sqrt{1-e_1^2}}\dot{e}_1
+\frac{m'_2\sqrt{1-e_2^2}}{2\sqrt{a_2}}\dot{a}_2-\frac{m'_2\sqrt{a_2}e_2}{\sqrt{1-e_2^2}}\dot{e}_2.
\end{equation}

The total angular momentum of the whole system (including the three-body system under study and the external component) is conserved during migration. Thus, we can write that $L_{\rm{orb}}+L_{\rm{ext}}=const$, where the external component of the total angular momentum is denoted as $L_{\rm{ext}}$.  It should be emphasized that the generic definition ``ext" does not necessarily mean that the exchange of the orbital angular momentum occurs with an exterior medium.  For instance, in the case of tidal interactions of planets with a central star, the `external' contribution to the angular momentum comes from the rotation of the star, as described in the next section. On the other hand, in the case of disc-planet interactions, there is a flux of angular momentum between the planet system and an external protoplanetary disc, as described in Section \ref{sec-3}.

For sake of simplicity, we restrict our investigation to the case in which only one planet is directly affected by dissipative forces and, consequently, its semi-major axis is changed. This implies that the semi-major axis of the other planet remains unchanged during secular evolution of the system. Assuming that the $i\,^{th}$-\,body is affected by dissipative forces and combining Equations (\ref{aidot}) -- (\ref{Lorbdot}), we use the conservation of the total angular momentum to find that
\begin{equation}\label{e1esecdot}
\dot{e}_i\,^{\rm{sec}}=-\frac{m'_j}{m'_i}\sqrt{\frac{a_j}{a_i}}
\frac{e_j}{e_i}\sqrt{\frac{1-e_i^2}{1-e_j^2}}\,\dot{e}_j\,^{\rm{sec}}
\end{equation}
and
\begin{equation}\label{e1extdot}
\dot{e}_i\,^{\rm{ext}}=\frac{(1-e_i^2)}{2a_ie_i}
\dot{a}_i+\frac{\sqrt{1-e_i^2}}{m'_i\sqrt{a_i}e_i}\dot{L}_{\rm{ext}},
\end{equation}
where $i\neq j$ and $e_i\neq0$. Equation (\ref{e1esecdot}) shows that, under mutual secular perturbations, the eccentricities of the planetary orbits oscillate in anti-phase. Equation (\ref{e1esecdot}) also shows that the evolution of $e_j$ depends implicitly on the migrating evolution of the companion, through the external variations of $a_i$ and $e_i$.

Equation (\ref{e1extdot}) can be re-written as
\begin{equation}\label{e1extdot2}
\dot{e}_i\,^{\rm{ext}}=(1-\alpha)\frac{(1-e_i^2)}{2a_ie_i}\dot{a}_i,\qquad e_i\neq0,
\end{equation}
where
\begin{equation}\label{alpha0}
\alpha=-\frac{2a_i}{\dot{a}_i}\frac{\dot{L}_{\rm{ext}}}{L_i},\qquad \dot{a}_i\neq0,
\end{equation}
with $L_i$ being the partial angular momentum of the body affected by dissipation. The function $\alpha$ can thus be used as a convenient measure of the non-conservation of $L_{\rm{orb}}$, since it quantifies the variation of the external component of the orbital angular momentum of the system. Indeed, if $\alpha=0$, $L_{\rm{ext}}$ remains constant and $L_{\rm{orb}}$ is conserved during migration. According to Equation (\ref{e1extdot2}), the limit case $e_i=0$ implies that $\dot{a}_i=0$ for $\alpha=0$, that is, the migration must be stopped when the planet orbit is circularized in the system with no exchange of angular momentum.




For $\alpha \neq 0$, the portion $1-\alpha$ of the $L_{\rm{orb}}$-\,change produced by $\dot{a}_i$ is absorbed by the system through the variation of $e_i$, while the rest ($\alpha$) is transferred to $L_{\rm{ext}}$. We say that there is a leakage (or gain) of the orbital angular momentum in the system. According to Equation (\ref{e1extdot2}), for $\alpha=1$, the dissipative force produces no change of the eccentricity and the change in $L_{\rm{orb}}$ due to $\dot{a}_i$ is totally transferred to $L_{\rm{ext}}$. This particular case of the angular momentum exchange was used to modeling of planet-planetesimal interactions (Malhotra 1995).

The range of theoretical values of $\alpha$ is large; according to Equation (\ref{alpha0}), $\alpha$ can be positive or negative, depending on the signs of $\dot{a}_i$ and $\dot{L}_{\rm{ext}}$. Moreover, $\alpha$ is a function of the orbital elements of the migrating planet and, consequently, varies during migration. Knowing that each physical process is characterized by a specific law of angular momentum exchange, in the next sections, we develop the $\alpha$-function for several kinds of dissipative interactions.  Using Equation (\ref{e1extdot2}), we write $\alpha$ in general form as
\begin{equation}\label{alpha}
\alpha=1-{\cal{F}}(e_i, \textrm{\textbf{par}})\frac{2e_i^2}{1-e_i^2},
\end{equation}
where ${\cal{F}}(e_i,\textrm{\textbf{par}})$ is a characteristic function of the migration process, defined through the condition
\begin{equation}\label{F}
\frac{\dot{e}_i^{\rm{ext}}}{e_i}={\cal F}(e_i, \textrm{\textbf{par}})\frac{\dot{a}_i}{a_i},
\end{equation}
where the vector $\textrm{\textbf{par}}$ is composed of physical parameters of the process under study.

\subsection{Evolutionary routes in the phase space}\label{sec-1-2}

As shown in Michtchenko and Rodr\'iguez (2011), under assumption that
dissipation is weak and slow, the evolutionary routes of the migrating non resonant planets
are traced by the Mode I and Mode II stationary solutions of the conservative
secular problem (see also Hadjidemetriou and Voyatzis 2010).
The ultimate convergence and the evolution of the system
along one of these secular modes of motion are determined by the
condition that the dissipation rate is smaller than the proper
secular frequency of the system. We have shown that, for values of $\alpha$
less than 1, the Mode I secular solution (characterized by aligned orbits) plays a role
of an attractive center when planetary orbits diverge and the condition $\sqrt{a_1/a_2} <
m_2/m_1$ is satisfied, or, when planetary orbits converge and $\sqrt{a_1/a_2} > m_2/m_1$.
In opposite cases, the Mode II (characterized by anti-aligned orbits) is
the attractive center of the migrating two-planet system.

In practice, the calculation of evolutionary tracks is simple. The current
location of the system in the phase space provides the starting values of the
orbital angular momentum, $L_{\rm{orb}}$, and the semi-major axis $a_i$ of the planet
directly affected by dissipation. The set, composed by $L_{\rm{orb}}$, masses and semi-major axes, uniquely defines
two stationary solutions ($e_1^{\ast},e_2^{\ast}$) of the conservative secular
problem at the current configuration. In order to model migration process, the semi-major axes of the affected planet is incremented by $\Delta a_i$ (which can be either positive or
negative), while the semi-major axis of the other planet is kept unaltered.
The orbital angular momentum of the system is then corrected by the amount
$$
\Delta L_{\rm{orb}} = \alpha \frac{L_i}{2a_i}\Delta a_i,
$$
where the function $\alpha$ is defined by Equation (\ref{alpha}). For a new set of $a_i$ and $L_{\rm{orb}}$ values (with fixed masses), we calculate new
values of stationary solutions, ($e_1^{\ast},e_2^{\ast}$). The
procedure is repeated until the system reaches domains with no possible
stationary solutions. The obtained values of ($e_1^{\ast},e_2^{\ast}$) are
then plotted on the representative planes ($n_1/n_2$, $e_i$), where $n_i$
($i=1,2$) are mean motions of the planets. The family of stationary solutions shows two evolutionary routes, one of which the system will follow during migration.

In the described above process, stationary solutions can be calculated using
the precise semi-analytical approach developed in Michtchenko and Malhotra
(2004), with no restrictions on the values of planet eccentricities. An
alternative is the use of the expansion of the disturbing function in series
of $a_1/a_2$, $e_1$ and $e_2$, given by
$$
{\cal
R}=\sum_{l=0}^N\sum_{m=0}^l\sum_{l_1',l_2'=0}^{N'}R_{l,m,l_1',l_2'}\Bigg{(}\frac{a_1}{a_2}\Bigg{)}^le_1^{l_1'}e_2^{l_2'}\cos(m\Delta\varpi),
$$
where $R_{l,m,l_1',l_2'}$ are numerical coefficients which do not depend on
the physical and orbital parameters of the system and
$\Delta\varpi\equiv\varpi_1-\varpi_2$ is the difference of planetary
longitudes of pericenter.

The main difficulty in the calculation of migration
routes is that the characteristic function ${\cal{F}}(e_i,
\textrm{\textbf{par}})$, which is needed to obtain $\alpha$ through Equation
(\ref{alpha}), is unknown a priory. Thus, the next sections of this paper will
be devoted to calculate the function ${\cal{F}}$ for some specific dissipative processes.

\section{Tidal interactions}\label{sec-2}

We first consider the case of tidal interactions, which affect orbital elements and rotations of the deformed bodies, while energy is dissipated due to internal friction. The tidal force is inversely proportional to the $7\,^{th}$ power of the distance between interacting bodies and is effective in both planetary and satellite systems (e.g. Mignard 1979). In this paper, we assume that only the central and close-in bodies are mutually affected by the raised tides, whereas the outer companion is not directly affected by the tidal force. The orbital planes of planets are assumed to be coincident.

\subsection{A system with a close-in planet}\label{sec-2-1}

We consider a short-period planet orbiting a slow-rotating central star ($\Omega_0\ll n_1$), where $\Omega_0$ and $n_1$ are the angular velocity of rotation of the star and the mean orbital motion of the planet, respectively.

The long-term variations (averaged over the inner planet orbital period) of the planetary elements, due to the combined effects of tides raised on the star and on the inner planet, are written, up to ${\cal O}(e_1^3)$, as:
\begin{equation}\label{adot_cumul}
<\dot{a}_1>\,=-\frac{4}{3}n_1a_1^{-4}\hat{s}[(1+23e_1^2)+7e_1^2D]
\end{equation}
and
\begin{equation}\label{edot_cumul}
<\dot{e}_1\,^{\rm{ext}}>\,=-\frac{2}{3}n_1e_1a_1^{-5}\hat{s}[9+7D],
\end{equation}
where
\begin{equation}\label{eq:D}
 D\equiv\hat{p}/2\hat{s},
\end{equation}
with
\begin{eqnarray}\label{ps}
\hat{p}\equiv\frac{9}{2}\frac{k_1}{Q_1}\frac{m_0}{m_1}R_1^5\quad{\rm and}\quad
\hat{s}\equiv\frac{9}{4}\frac{k_0}{Q_0}\frac{m_1}{m_0}R_0^5
\end{eqnarray}
being the strengths of stellar and planetary tides, respectively (Dobbs-Dixon et al. 2004; Ferraz-Mello et al. 2008, Rodr\'iguez and Ferraz-Mello 2010). $k_i$ and $Q_i$ are the Love number and the dissipation function of the deformed body, where the indices $i=0$ and $1$ stand for the central and inner bodies, respectively. In above equations, it is assumed a quasi-synchronous rotation state of the close-in planet ($\Omega_1\simeq n_1$). In addition, is is also assumed a linear dependence between phase lags and the corresponding frequencies appearing in the decomposition of the tidal potential of each body (Mignard 1979; Ferraz-Mello et al. 2008). For sake of simplicity, we consider that $Q_i$ are constant; however, it should be kept in mind that their dependence on frequencies can be important in the long-term evolution. The reader is referred to Efroimsky and Williams (2009) for more discussions on the frequency dependence of $Q_i$.

Combining Equations (\ref{F}) and (\ref{adot_cumul})--(\ref{edot_cumul}), we obtain the characteristic function
\begin{equation}\label{F2}
{\cal F}(e_1,D)=\frac{9+7D}{2\left[1+(23+7D)e_1^2\right]},
\end{equation}
where $\textrm{\textbf{par}}$ is defined by the parameter $D$. The $L_{\rm{orb}}$--leakage or the function $\alpha$ is then calculated through Equation (\ref{alpha}) to give

\begin{equation}\label{alfaF2}
\alpha=1-\frac{(9+7D)e_1^2}{(1-e_1^2)\left[1+(23+7D)e_1^2\right]}.
\end{equation}

\begin{center}
\begin{figure}
\includegraphics[width=0.5\linewidth,angle=270]{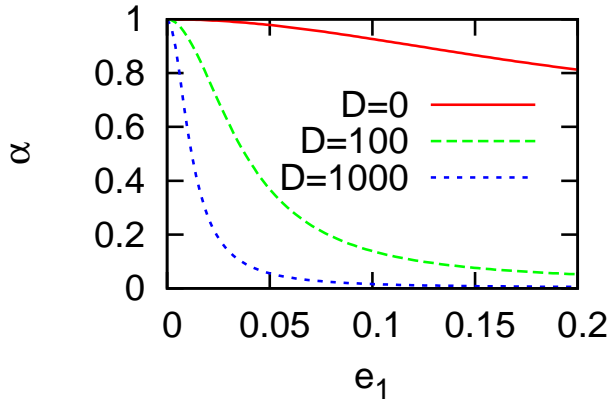}
\caption{\small The $L_{\rm{orb}}$--leakage $\alpha$ as a function of $e_1$, in the case of a close-in planet tidal evolution. The curves are parameterized by several constant values of $D$.}
\label{alpha-fig}
\end{figure}
\end{center}

Figure \ref{alpha-fig} shows the variation of $\alpha$ with $e_1$. We observe that, for all values of $D$, $\alpha$ varies in the range between 0 and 1. The immediate consequence is that the tidal decay of a close-in planet is generally accompanied by the loss of orbital angular momentum of the system. The portion $1-\alpha$ is absorbed by the system, damping the eccentricity of the inner planet. The amount $\alpha$ is transferred to the central star and accelerates the stellar rotation (see discussion of the next paragraph).

The stellar rotation changes due to tides on the star and, from the above discussion, it is clear that its variation plays a role of external source of the angular momentum. The rate of variation of the rotational angular momentum is given by $\dot{L}_{\rm{ext}}=\dot{L}_{\rm{rot}}=C_0\dot{\Omega}_0$, where $C_0$ is the stellar moment of inertia (assuming that the inner planet has reached stationary rotation, we can neglect the spin variation of the planet; e.g. Rodr\'iguez et al. 2011 for more details). Hence, using the definition of $\alpha$ given in Equation (\ref{alpha0}), we obtain, for $\dot{a}_1\neq0$,
\begin{equation}\label{alpha-tide}
\alpha=-\frac{2C_0a_1}{L_1\dot{a}_1}\dot{\Omega}_0.
\end{equation}
Since $\dot{a}_1<0$ and $\alpha>0$ during orbital decay of the inner planet, above equation results in $\dot{\Omega}_0>0$, as previously discussed.

On one hand, Equation (\ref{alpha-tide}) shows that, for $\alpha=0$, $\Omega_0$ is a constant and there is no tides on the star (i.e. $\hat{s}=0$). In this case, the orbital angular momentum of the system is conserved. Inversely, knowing that $D\equiv\hat{p}/2\hat{s}$, $\hat{s}=0$ means that $D\rightarrow\infty$ and, according to Equation (\ref{F2}), ${\cal F}(e_1,D)=0.5e_1^{-2}$. Hence, to second order in $e_1$ and according to Equation (\ref{alpha}), we have $\alpha=0$. Note that, for $\alpha\simeq0$ (very large $D$), the migration is halted when the orbit circularizes, that is, $a_1$ becomes constant because there is no orbital angular momentum transfer and $e_1=0$.

On the other hand, when planetary tides are neglected, we have $\hat{p}=D=0$ and, up to second order in $e_1$, $(1-\alpha)=9e_1^2/(1+23e_1^2)$. This last expression shows that, after circularization of the inner planet orbit ($e_1=0$), the orbital angular momentum change due to the orbital decay is totally transferred to the stellar spin ($\alpha=1$). Evidence for excess of rotational angular momentum of the parent stars due to the tidal interaction with close-in giant planets have been already addressed in previous studies (e.g., Pont 2009).

It is important to note that, during tidal interactions, $\alpha$ is not constant but varies as a function of $e_1$, according to Equation (\ref{alfaF2}). The sign of $\alpha$ is always positive, meaning that the decreasing of the orbital angular momentum due to the orbital decay of the inner planet is compensated by both the damping of $e_1$ and the increase of $\Omega_0$.
\begin{center}
\begin{figure}
\includegraphics[width=0.7\linewidth,angle=0]{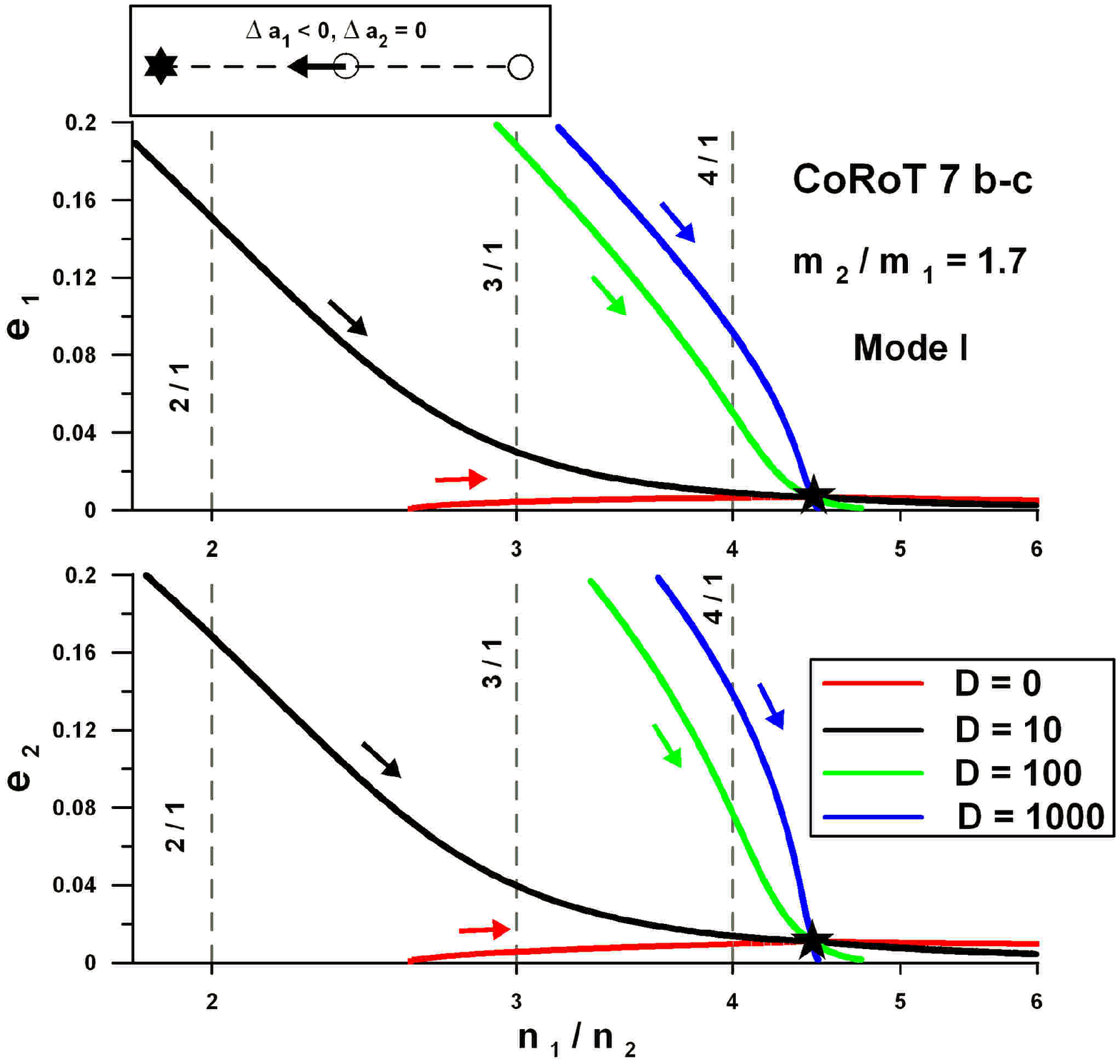}
\caption{\small Evolutionary routes of CoRoT-7 system parameterized by constant values of $D$. The angular momentum exchange is determined by the $\alpha$--function (\ref{alpha}) defined through (\ref{alfaF2}). The current position of the system is shown by a star symbol, whereas the locations of some mean-motion resonances are indicated by dashed lines. A schematic view of the migration scenario is shown in the box illustration at the top panel.}
\label{figure1}
\end{figure}
\end{center}

Figure \ref{figure1} shows the evolutionary tracks of a migrating pair of planets. They
were obtained  as discussed in Section \ref{sec-1-2}, where the orbital angular
momentum variation for each curve was defined by the function $\alpha$, according to
Equation (\ref{alfaF2}). An application to the CoRoT-7 planetary system,
characterized by very small current eccentricities (Ferraz-Mello et al. 2011)
was considered, whose results are plotted in the planes ($n_1/n_2$, $e_i$).
The adopted migration scenario is schematically shown on the top of the
figure. The divergent migration, together with the condition $m_2/m_1>\sqrt{a_1/a_2}$, results  in  the Mode I ($\Delta\varpi=0$) of secular motion as an attractive center (Michtchenko and Rodr\'iguez 2011). The current position of
the system is marked with a star symbol and the arrows  indicate  the
direction  of  the  evolution  (orbital  decay  of  the  inner planet).

We can note that migration tracks are sensible to the values of the parameter
$D$, and consequently, to the orbital angular momentum exchange.  For small
$D$ ($\alpha\simeq1$), corresponding to the nearly-total loss of the
angular momentum variation produced by the orbital decay of the inner planet,
both eccentricities vary only slightly during evolution. In this case, when
the strength of the stellar tide is dominating, the orbital configuration of the system in the past would be not different from the present one.

The situation is very different in the case of large values of $D$, when the
strength planetary tide is dominating. Large $D$-values imply that the
orbital angular momentum of the system is almost conserved ($\alpha\simeq0$).
In this case, the orbital configuration of the system in the past should be
characterized by high eccentricities, as higher as larger are $D$-values.
Therefore, in the cases with dominating planetary tides, the origin
of such high eccentric planetary orbits must be investigated.

Note that, during migration, the two-planet system could cross several mean-motion resonances, resulting in the temporary excitation of eccentricities but without trapping, since the orbits are divergent. Several numerical simulations performed in Rodr\'iguez et al. (2011) and Michtchenko and Rodr\'iguez (2011), show that the system ultimately returns to the stationary secular solution after leaving the mean-motion resonance.


\subsection{A system of satellites}\label{sec-2-2}

We consider now the case of a pair of satellites orbiting a rapidly rotating parent planet ($\Omega_0\gg n_1$). We assume that only the inner satellite is affected by tidal interactions with the planet. According to the linear tidal model, the averaged variations of the orbital elements are given by
\begin{equation}\label{adotsat}\nonumber
<\dot{a}_1>\,=\frac{4}{3}n_1a_1^{-4}\hat{s}[(1+(27/2-7D)e_1^2]
\end{equation}
and
\begin{equation}\label{edotsat}\nonumber
<\dot{e}_1^{\rm{ext}}>\,=\frac{2}{3}n_1e_1a_1^{-5}\hat{s}(11/2-7D)
\end{equation}
(Goldreich and Soter 1966; Ferraz-Mello et al. 2008). Note that the only difference with respect to the previous case is in the tides raised on the planet (or central body). We still note that the mean variations can be positive or negative, depending on the value of $D$. Hence, the combined effect of planet and satellite tides can induce either inward or outward migration, as well as either damping or excitation of eccentricity. The sign of $\dot{e}_1^{\rm{ext}}$ depends on the value of $D$ in such a way that $\dot{e}_1^{\rm{ext}}>0$ if $D < 11/14$. The sign of $\dot{a}_1$ depends on both $D$ and $e_1$-values. For small eccentricities ($e_1\le0.2$) and moderate $D$ (in the range from 1 to 5), we would expect outward migration and damping of eccentricity, a typical behavior observed in the tidal evolution of Solar System satellites (e.g. Peale 1999). In this case (i.e., $\dot{a}_1>0$ and $\dot{e}_1^{\rm{ext}} < 0$), the total angular momentum conservation implies that $\alpha > 1$ (see Equation (\ref{e1extdot2})) and, according to Equation (\ref{alpha-tide}), $\dot{\Omega}_0<0$. The Earth-Moon tidal evolution, with the satellite moving away from the Earth and the increasing length of day of the planet, corresponds to this scenario.
\begin{center}
\begin{figure}
\includegraphics[width=0.5\linewidth,angle=270]{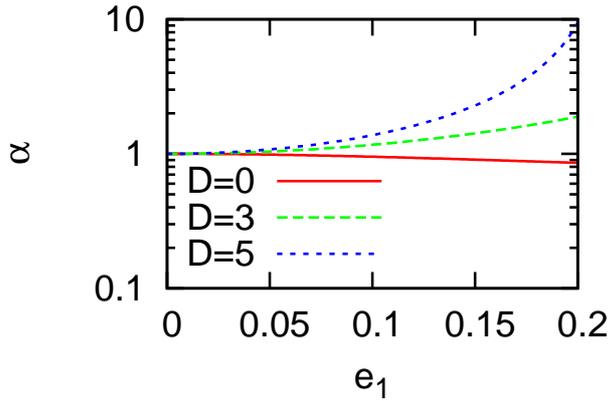}
\caption{\small Dependence of $\alpha$ on $e_1$, in logarithmic scale, for the tidal evolution of a close-in satellite. Illustrations are shown for different values of $D$. Here, $\alpha$ can adopt values larger than unity, but not in the close-in tidal evolution case (see Figure. \ref{alpha-fig} for comparison).}
\label{alpha2-fig}
\end{figure}
\end{center}

Combining Equations (\ref{alpha})-(\ref{F}) and (\ref{adotsat})-(\ref{edotsat}), we have
\begin{equation}\label{F3}
{\cal F}(e_1,D)=\frac{11/2-7D}{2[1+(27/2-7D)e_1^2]},
\end{equation}
and
\begin{equation}\label{alfaF3}
\alpha=1-\frac{(11/2-7D)e_1^2}{(1-e_1^2)[1+(27/2-7D)e_1^2]}.
\end{equation}
Figure \ref{alpha2-fig} shows the variation of $\alpha$ with $e_1$, for different values of $D$. With the exception of the example with $D=0$, $\alpha > 1$ always, indicating that the external source of angular momentum (planet rotation) injects angular momentum in the satellite system, which is used to expand the inner planet orbit and circularize both inner and outer orbits.
\begin{center}
\begin{figure}
\includegraphics[width=0.7\linewidth,angle=0]{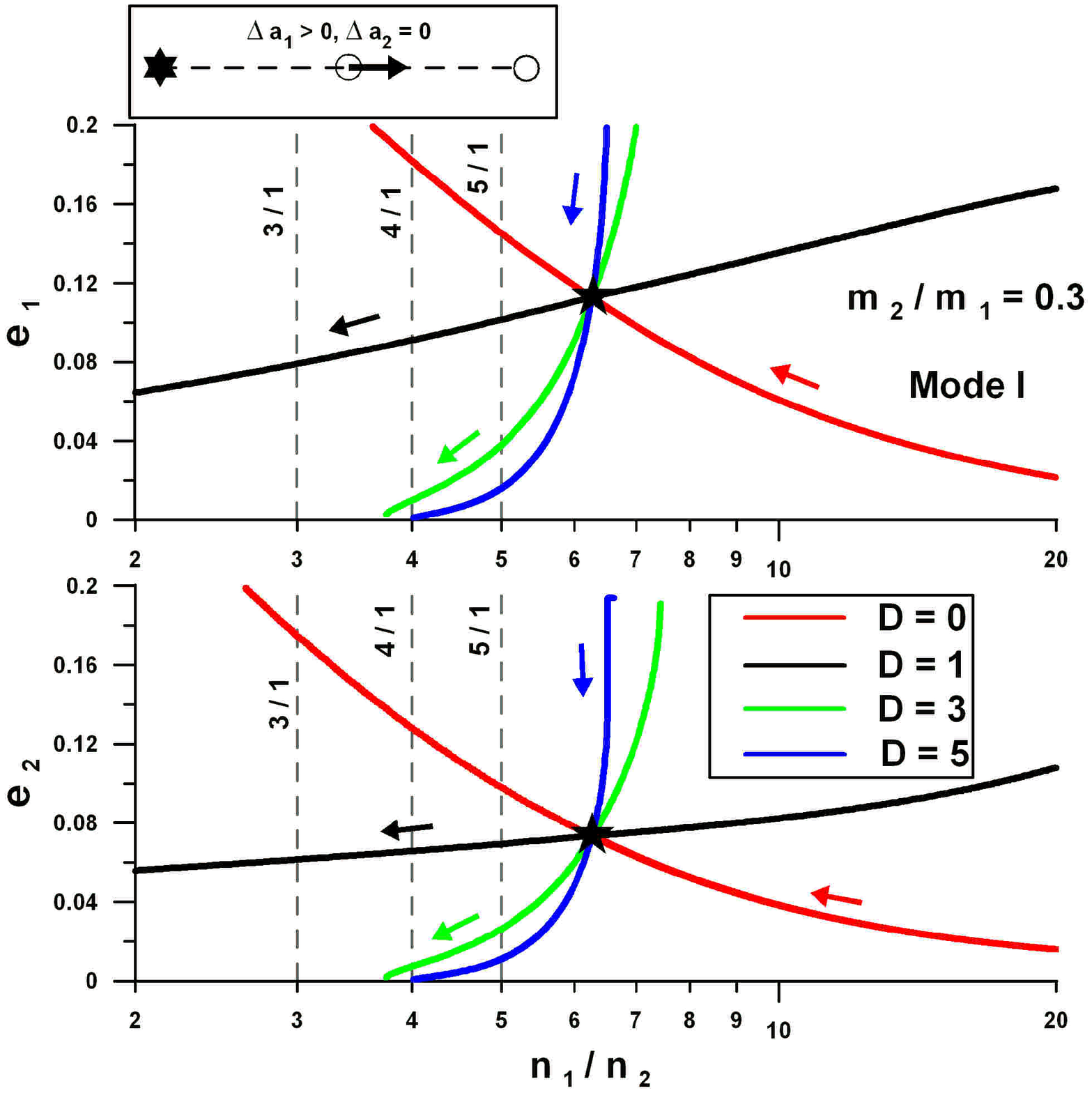}
\caption{\small The same as in Figure \ref{figure1}, except the angular momentum exchange is determined by the $\alpha$--function defined through (\ref{alfaF3}).}
\label{figure2}
\end{figure}
\end{center}

Figure. \ref{figure2} shows the migration routes of a hypothetical system in which the central body is an Earth-like planet, the inner satellite has the Moon mass and  the  outer  companion  is  three  times  smaller  than  the  Moon. The curves are parameterized by several values of $D$. For $D$ in the range from 0 to 5, the inner satellite migrates outward when the eccentricity of its orbit is not too large ($e_1<0.2$, according to Equation (\ref{adotsat})). Since the semi-major axis of the outer satellite is unaltered during the secular interaction with its companion, we have a convergent migration and, together with the condition $m_2/m_1<\sqrt{a_1/a_2}$, results in the Mode I ($\Delta\varpi=0$) as the attractive center. The direction of migration along evolutionary routes is shown by arrows in Figure. \ref{figure2}.

For $D\geq1$, the satellite eccentricities decrease during migration: larger is the value of $D$, more rapid is the damping of eccentricities. In this way, the satellite system approaches to main mean-motion resonances with low-eccentricity orbits which allow a smooth capture into one of these resonances (e.g. Tittemore and Wisdom 1988). The behavior of the system described by $D=0$ is different, because in this case the eccentricities of the satellites increase during migration (see Equation (\ref{edotsat})). As a consequence, the capture inside a low-order mean-motion resonance would be strongly improbable.

\section{Disk-planet interactions}\label{sec-3}

In this section we study the migration of a two-planet coplanar system assuming that dissipative forces affect only the outer planet. We also assume that the two planets are interacting secularly, that is, the planets are far enough from any mean-motion resonance. Disc-planet interactions are the natural mechanism which results in the orbital decay of the outer planet due to energy and angular momentum exchange with an outer gaseous disc (e.g. Kley 2000, 2003).

\subsection{Evolutionary tracks under Stokes interactions}\label{sec-3-1}

For sake of simplicity, we suppose that the outer planet is forced to migrate under the action of a dissipative drag force (Stokes-like force) of the type

\begin{equation}\label{stokes}
\vec{f}=-10^{-\nu}(\vec{v}_2-\gamma\,\vec{v}_{2c}),
\end{equation}
where $\nu>0$ and $\vec{v}_{2c}$ is the Keplerian circular velocity at the astrocentric distance $r_2$. Due to the negative radial pressure gradient, the gas velocity is a bit less than the circular velocity at the same point. Consequently, the value of the factor $\gamma$ should be slightly smaller than unity (see Adachi et al. 1976; Patterson 1987). $\gamma$ is also frequently used as a free parameter of the problem, in order to model different conditions of disc-planet interactions (e.g. Beaug\'e et al. 2006). Several previous works have investigated the evolution of a two-planet system evolving under a Stokes drag force (e.g., Beaug\'e at al. 2006; Hadjidemetriou and Voyatzis 2010), with the majority devoted to the study of the evolution inside mean-motion resonances.

In order to obtain the variations of the semi-major axis and the eccentricity of the outer planet produced by the force $\vec{f}$, we use the Euler-Gauss's equations (Brouwer and Clemence 1961):
\begin{eqnarray}\label{gauss}
\frac{da_2}{dt}&=&\frac{2}{n_2\sqrt{1-e_2^2}}\left[f_r\, e_2\, \sin u_2  + f_t(1+e_2\cos u_2) \right]\\
\frac{de_2}{dt}&=&\frac{\sqrt{1-e_2^2}}{n_2\, a_2}\left[f_r \sin u_2 + \frac{f_t}{e_2}\left(1+e_2\cos u_2-\frac{r_2}{a_2} \right)  \right]\nonumber\\
\frac{d\varpi_2}{dt}&=&\frac{\sqrt{1-e_2^2}}{n_2\, a_2\,e_2}\left[-f_r \cos u_2 + f_t\left(\frac{r_2}{a_2(1-e_2^2)}+ 1 \right)\sin u_2  \right]\nonumber\\
\frac{d\,l^0_2}{dt}&=&\frac{\sqrt{1-e_2^2}}{e_2}\left[(a_2(1-e_2^2)\cos u_2-2e_2r_2)f_r-(r_2+a_2(1-e_2^2))\sin u_2\,f_t\right],\nonumber
\end{eqnarray}
where $l^0_2$ is related to the mean anomaly, $l_2$, through $l_2=l^0_2+\int n_2dt$.
$f_r$ and $f_t$ are the radial and transverse components of $\vec{f}$, whereas $u_2$ is the true anomaly of the outer planet. The first-order averaging over the outer planet orbital period gives, to fourth order in $e_2$ and assuming $m_2\ll m_0$:
\begin{eqnarray}
<\dot{a}_2>&=&-10^{-\nu}a_2[2(1-\gamma)+\gamma(5e_2^2/8+119e_2^4/512)]\label{gaussa-medias}\\
<\dot{e}_2>&=&-10^{-\nu}\gamma\,e_2(1-13e_2^2/32)\label{gausse-medias}\\
<\dot{\varpi_2}>&=&0
\end{eqnarray}
(for comparison, see Beaug\'e and Ferraz-Mello 1993). It follows from above equations that the disc-planet interactions produce orbital decay and circularization. If we set $\gamma=1$ (see Hadjidemetriou and Voyatzis 2010) and $e_2=0$ into Equation (\ref{gaussa-medias}), then we have
$\dot{a}_2=0$, indicating that there is no orbital decay of the outer planet in the case of circular orbit.

Using Equations (\ref{alpha})-(\ref{F}) and (\ref{gaussa-medias})-(\ref{gausse-medias}), we obtain
\begin{equation}\label{alpha3}
{\cal F}(e_2,\gamma)=\frac{\gamma\,(1-13e_2^2/32)}{2(1-\gamma)+\gamma(5e_2^2/8+119e_2^4/512)}
\end{equation}
and
\begin{equation}\label{alfaF4}
\alpha=1-\frac{2\gamma\,(1-13e_2^2/32)e_2^2}{(1-e_2^2)\left[2(1-\gamma)+\gamma(5e_2^2/8+119e_2^4/512)\right]},
\end{equation}
where $\textrm{\textbf{par}}$ is given by the parameter $\gamma$.
\begin{center}
\begin{figure}
\includegraphics[width=0.5\linewidth,angle=270]{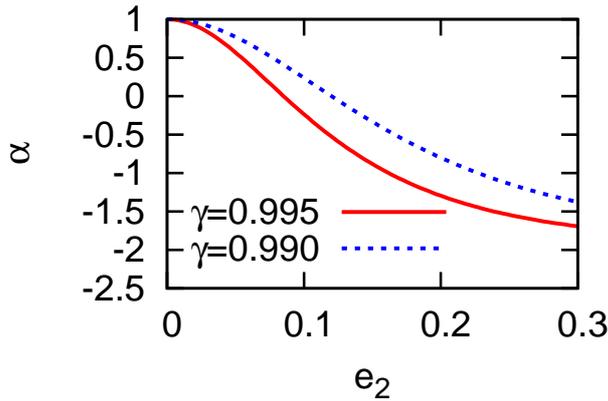}
\caption{\small Variation of $\alpha$ with $e_2$, for two values of $\gamma$, corresponding to the case of disc-planet interaction simulated with a drag force of the type (\ref{stokes}). Note that, depending on $e_2$, $\alpha$ can adopt negative values (see text for details).}
\label{alfa-disco-fig}
\end{figure}
\end{center}
Figure \ref{alfa-disco-fig} shows the function $\alpha(e_2,\gamma)$ parameterized by two values of $\gamma$.
We note that, as in the case of star-planet tidal interactions (see Sect. \ref{sec-2-1}), $\alpha$ is always smaller than 1, which means that the orbital decay of the outer planet is always accompanied by damping of the eccentricity of its orbit (see Equation (\ref{e1extdot2})). We also see that, for $e_2=0$, $\alpha=1$ for any value of $\gamma$ (we have discussed that $\gamma<1$), indicating that, after circularization of the planet orbit, the change of $L_{\rm{orb}}$ due to the orbital decay is totally transferred to $L_{\rm{ext}}$, in this case, to the gaseous disc.

In contrast with the case of star-planet tidal interactions, the function $\alpha$ is not saturated at 0 (see Figure \ref{alpha-fig}), but decreases monotonously with eccentricities towards negative values. According to Equation (\ref{alpha0}), negative values of $\alpha$ imply in the loss of  angular momentum by the disc during the orbital decay of the outer planet ($\dot{L}_{\rm{ext}} < 0$). Thus, during migration, the pair of planets can either gain or lose orbital angular momentum, depending on the eccentricity of the migrating planet.

A transition between leakage or gain of angular momentum occurs at $\alpha=0$, when $e_2$ reaches a critical value $e_{2}^{\rm{cr}}$ which is determined from Equation (\ref{alfaF4}), for a given $\gamma$. For instance, for $\gamma=0.995$, $e_{2}^{\rm{cr}}\simeq0.085$. For $e_2=e_{2}^{\rm{cr}}$, the orbital angular momentum of the planet system is conserved (since $\alpha=0$), however, this condition is only transitory. For $e_2<e_{2}^{\rm{cr}}$, we have $0<\alpha<1$, indicating that a portion $\alpha$ of the orbital angular momentum change due to migration is removed from the system and transferred to the disc ($\dot{L}_{\rm{ext}}>0$, see Equation (\ref{alpha0})), while the portion $1-\alpha$ is absorbed by the system altering the value of $e_2$. When $e_2>e_{2}^{\rm{cr}}$, we have $\alpha<0$ and the system gains angular momentum from the disc. Note that, in the latter case, the planet system gains angular momentum, despite the semi-major axis of the outer planet is decreasing. 

\begin{center}
\begin{figure}
\includegraphics[width=0.7\linewidth,angle=0]{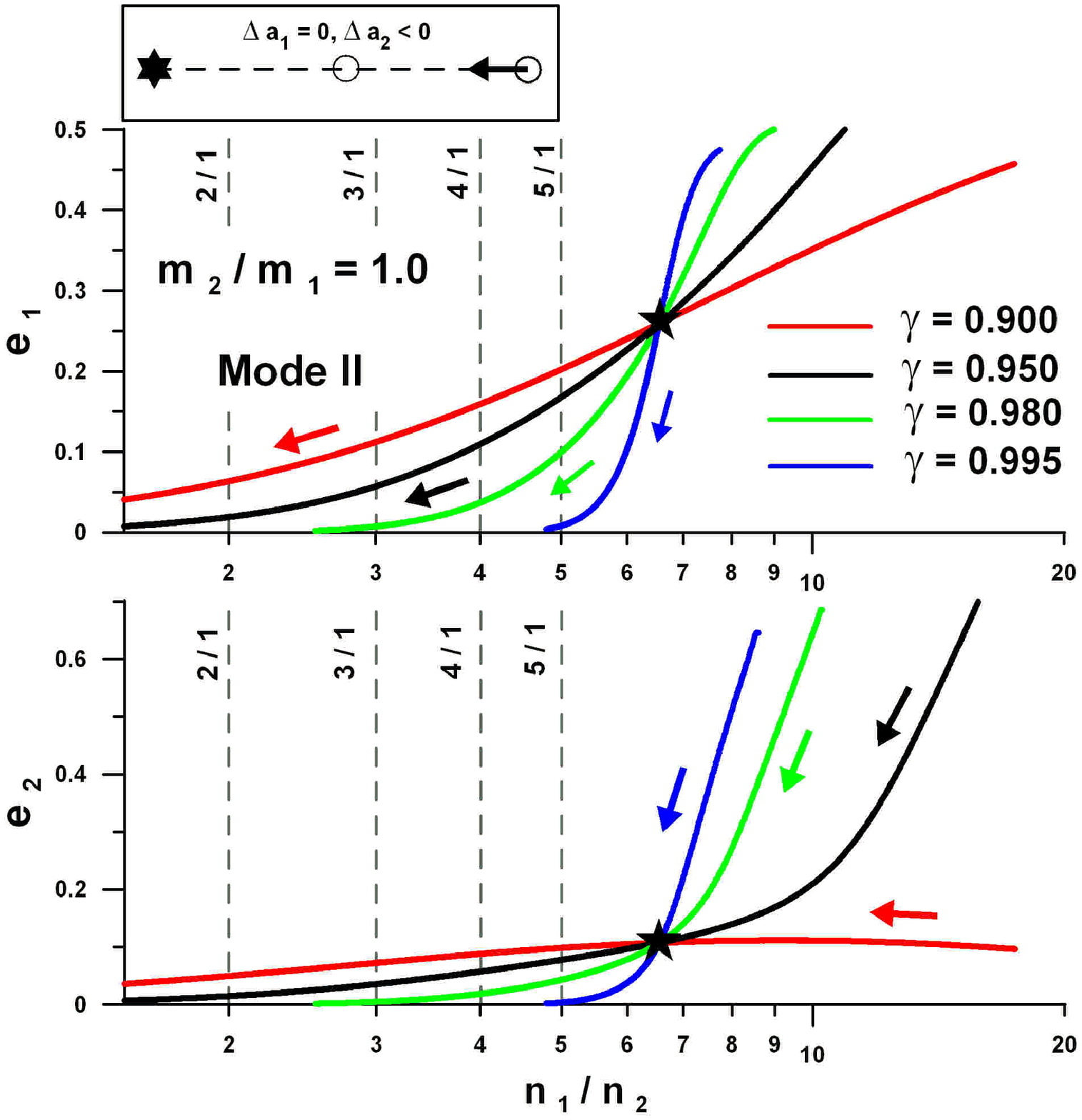}
\caption{\small Evolutionary routes for a disc-planet interaction simulated through the force given in Equation (\ref{stokes}), parameterized by constant values of $\gamma$. In this case, the angular momentum exchange is determined by the $\alpha$--function (\ref{alfaF4}).}
\label{figure3}
\end{figure}
\end{center}

The evolutionary routes of the planet pair, obtained for several values of the parameter $\gamma$ and for the angular momentum exchange defined by (\ref{alfaF4}), are shown in Figure \ref{figure3}. The hypothetical system is composed of a Sun-like central body and two planets of equal mass. The migration scenario shown on the top of the figure corresponds to the outer planet moving towards the inner planet. In this case, migration is convergent and the Mode II plays a role of an attractive center, since $m_2/m_1>\sqrt{a_1/a_2}$ (see Michtchenko and Rodr\'iguez 2011). Planet eccentricities are damped during migration (see Equation (\ref{gausse-medias})), while the system approaches to one of the main mean-motion resonances, whose locations are indicated in Figure \ref{figure3}. Because of small eccentricities, the probability of capture in a mean-motion resonance would be high, mainly for larger $\gamma$-values. It  is  worth  noting  that,  to  achieve  the  current  position  of  the  system (marked by a star symbol), a past orbital configuration with higher eccentricities is needed for all considered values of $\gamma$.

It is important to stress that, during the migration path, the system crosses several mean-motion resonances, as shown in Figure \ref{figure3}. One should keep in mind that, when the rate of dissipation is small, the system may be trapped one of such resonances. Since we restrict our investigations to the secular evolution of the systems, we fix the initial configurations of planet pairs far away from mean-motion resonances and, according to our model, neglect the possibility of a resonant capture.

\subsection{The factor $K$}\label{sec-3-2}

Several works have simulated the disc-planet interaction adopting a simplified model of planetary migration (e.g. Lee and Peale 2002; Beaug\'e et al. 2006, among others). The method avoids the use of hydrodynamic numerical simulations involving an accurate description of the interaction process between the gas and the planet body. The model introduces the relationship between the migration rates of semi-major axis and eccentricity as follows

\begin{equation}\label{K1}
\frac{\dot{e}_2}{e_2}=-K\left|\frac{\dot{a}_2}{a_2}\right|,
\end{equation}
where $K$ is a constant (positive) parameter. Given a value of $K$, the migrating systems are modeled through numerical integrations of the equations of planetary motion, with constant perturbations in the orbital elements according to Equation (\ref{K1}).

As shown in Beaug\'e and Ferraz-Mello (1993) and Gomes (1995), the time evolution of orbital elements due to the action of the drag force given by (\ref{stokes}) is, to first order in eccentricity:

$$a_2(t)=a_{20}\,\textrm{exp}(-At)\qquad \textrm{and}\qquad e_2(t)=e_{20}\,\textrm{exp}(-Et),$$
where $A$ and $E$ are the inverse of the e-folding times of $a_2$ and $e_2$, respectively. Introducing above equations into Equation (\ref{K1}) we obtain that $K=E/A$. Moreover, using Equations (\ref{gaussa-medias})-(\ref{gausse-medias}) we also obtain that, to first order in $e_2$, $K=\gamma/2(1-\gamma)$. In this way, we have shown that the condition (\ref{K1}), with a constant value of $K$, can be directly obtained from the averaged variations of the orbital elements produced by the Stokes force. It is worth noting that, for $\gamma=0.995$, which is the usually adopted value (Beaug\'e and Ferraz-Mello 1993), we obtain $K = 100$, which matches the value determined empirically by Lee and Peale (2002) for the GJ 876 resonant system (the planets b and c evolve inside the 2/1 mean-motion resonance).

\begin{center}
\begin{figure}
\includegraphics[width=0.5\linewidth,angle=270]{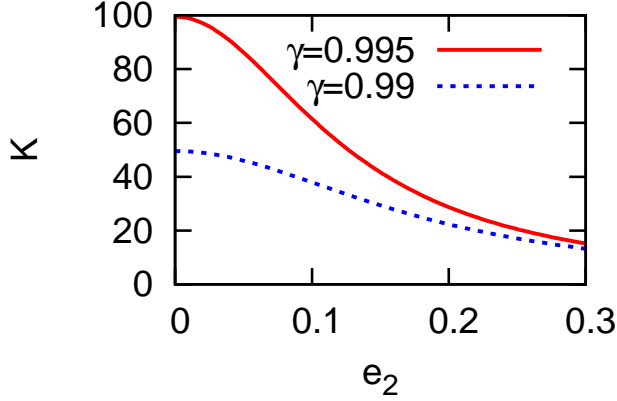}
\caption{\small Variation of $K$ with $e_2$, for two values of $\gamma$. When $e_2=0$ and $\gamma=0.995$, $K=100$ is obtained, reproducing the result of previous works (Lee and Peale 2002). Moreover, $K$ strongly varies with $e_2$ and for high eccentric orbits is almost independent on the value of $\gamma$.}
\label{k-alfa}
\end{figure}
\end{center}

The relationship (\ref{K1}) is frequently applied in moderate and even high eccentricity domains, specially when the evolution of resonant pairs of planets is modeled. However, comparing Equations (\ref{F}) and (\ref{K1}) we note that $K={\cal F}(e_2,\gamma)$, that is, the assumption of constant $K$ is not appropriate in the case of eccentric orbits and the function ${\cal F}(e_2,\gamma)$, given by Equation (\ref{alpha3}), should be used instead $K=const$ (see also Kley et al. 2004). The variation of $K$ as a function of $e_2$ is shown in Figure \ref{k-alfa}.

\subsection{A general drag force}\label{sec-3-3}

The Stokes force (\ref{stokes}) belongs to the class of a more general dissipative force given by
\begin{equation}\label{stokes-gral}
\vec{f}=-10^{-\nu}\rho(r_2)(\vec{v}_2-\gamma\,\vec{v}_{2c}),
\end{equation}
where $\rho(r_2)$ is the density profile of the gas in the disc (see Smart 1960). In this work, we assume that $\rho(r_2)=r_2^{-\beta}$, with $\beta\ge0$ (see Kominami and Ida 2002, for an example with $\beta=2$). Some works also adopt a more general dependence on the relative velocity, introducing powers of $(\vec{v}_2-\gamma\,\vec{v}_{2c})$ larger than 1. This case is related to high Reynolds numbers and correspond to regions of turbulence in the gas medium.

Using the Euler-Gauss equations with the external force (\ref{stokes-gral}) and applying the averaged procedure (over orbital periods), we obtain the long-term variations of semi-major axis and eccentricity of the outer planet, up to ${\cal O}(e_2^5)$, as follows
\begin{equation}\label{dota-gral}
<\dot{a}_2>\,=-10^{-\nu}a_2^{1-\beta}\,[2(1-\gamma)+e_2^2\,{\cal G}_1(\gamma,\beta)+e_2^4\,{\cal G}_2(\gamma,\beta)],
\end{equation}
\begin{equation}\label{dote-gral}
\hspace{-1.2cm}<\dot{e}_2>\,=-10^{-\nu}a_2^{-\beta}\,e_2\,[\gamma(1-\beta)+\beta+e_2^2\,{\cal G}_3(\gamma,\beta)],
\end{equation}
where
\begin{eqnarray}
{2\cal G}_1 = 3\beta+\beta^2+\gamma\left(\frac{5}{4}-2\beta-\beta^2\right),\nonumber
\end{eqnarray}
\begin{eqnarray}\label{G2}
16{\cal G}_2&& = 7\beta+\frac{23}{2}\beta^2+5\beta^3+\frac{1}{2}\beta^4,\nonumber\\
&&+\gamma\left(\frac{119}{32}-3\beta-\frac{27}{4}\beta^2-4\beta^3-\frac{1}{2}\beta^4\right),\nonumber
\end{eqnarray}
\begin{eqnarray}\label{G3}
4{\cal G}_3 = -3\beta+\frac{3}{2}\beta^2+\frac{1}{2}\beta^3+\gamma(\beta-1)\left(\frac{13}{8}-\beta-\frac{1}{2}\beta^2\right).\nonumber
\end{eqnarray}
It is clear that the case $\beta=0$ corresponds to the previously studied Stokes force (see Equations (\ref{gaussa-medias})-(\ref{gausse-medias})).

Interesting features can be highlighted from the above equations. On one hand, we note that fixing $a_2$ and $e_2$, the rate of migration increases when $\beta<1$ and decreases if $\beta>1$. Moreover, for $\beta=1$, $<\dot{a}_2>$ does not depends on $a_2$. On the other hand, the rate of $e_2$\,-\,damping decreases for $\beta>0$ and it is independent of $a_2$ for the Stokes drag ($\beta=0$).

\begin{center}
\begin{figure}
\includegraphics[width=0.5\linewidth,angle=270]{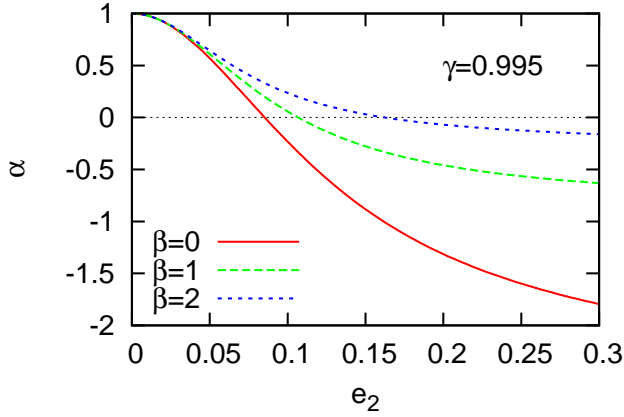}
\caption{\small Variation of $\alpha$ with $e_2$ for the case of a general dissipative force given by (\ref{stokes-gral}). Three values of the power $\beta$ are illustrated for $\gamma=0.995$.}
\label{k-alfa-gral}
\end{figure}
\end{center}

Through Equations (\ref{alpha})-(\ref{F}) and (\ref{dota-gral})-(\ref{dote-gral}), the characteristic function ${\cal F}(e_2,\gamma,\beta)$ is given by

\begin{equation}\label{F-gral}
{\cal F}(e_2,\gamma,\beta)=\frac{[\gamma(1-\beta)+\beta+e_2^2\,{\cal G}_3(\gamma,\beta)]}{[2(1-\gamma)+e_2^2\,{\cal G}_1(\gamma,\beta)+e_2^4\,{\cal G}_2(\gamma,\beta)]}
\end{equation}
and the function $\alpha$
\begin{equation}\label{alpha-gral}
\alpha=1-\frac{2e_2^2[\gamma(1-\beta)+\beta+e_2^2\,{\cal G}_3(\gamma,\beta)]}{(1-e_2^2)[2(1-\gamma)+e_2^2\,{\cal G}_1(\gamma,\beta)+e_2^4\,{\cal G}_2(\gamma,\beta)]},
\end{equation}
where \textbf{par} is given by two parameters of the system, $\gamma$ and $\beta$. Figure \ref{k-alfa-gral} shows the variation of $\alpha(e_2,\beta,\gamma)$ as a function of $e_2$, for $\gamma=0.995$. We show curves parameterized by $\beta=0,1,2$. Note that $\alpha$ has a strong dependence on $e_2$ for all $\beta$. For high eccentricity ($e_2\ge0.2$), $\alpha<0$ for all values of $\beta$ illustrated. When $\alpha<0$, the planetary system gains orbital angular momentum, because $\dot{a}_2<0$ and, looking at Equation (\ref{alpha0}), we have $\dot{L}_{\rm{ext}}<0$. Moreover, the gain is larger as smaller is the power $\beta$. In the limit of very small eccentricity ($e_2\le0.05$), $\alpha$ is almost independent on $\beta$.

\begin{center}
\begin{figure}
\includegraphics[width=0.7\linewidth,angle=0]{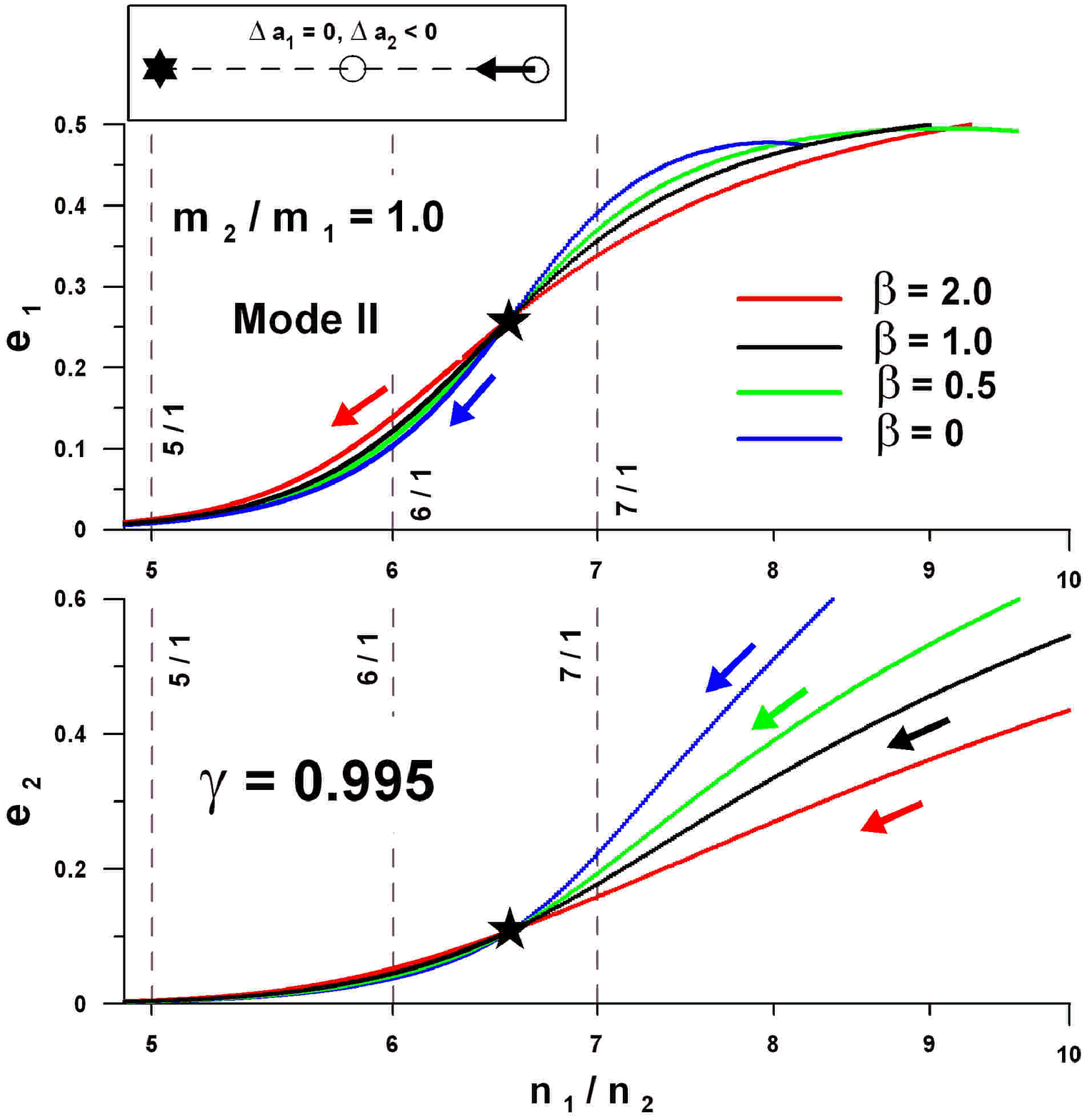}
\caption{\small Evolutionary routes for a disc-planet interaction simulated through the general force (\ref{stokes-gral}), parameterized by constant values of the power $\beta$ and $\gamma=0.995$. The angular momentum exchange is determined by the corresponding $\alpha$--function, given by Equation (\ref{alpha-gral})}
\label{figure4}
\end{figure}
\end{center}

The dependence of the planet evolution on the power $\beta$ is illustrated in Figure \ref{figure4}, where we show the migration paths of the  fictitious two-planet system, previously analyzed in Section. \ref{sec-3-1}. Several values of $\beta$ are used, with $\gamma$ fixed at 0.995. We can note that, at least for the given $\gamma$, the different density profile distributions produce only quantitative differences in the damping of eccentricities, which are significant only at high eccentricity domains. The convergent migration of the planets can results in the crossing or capture in a mean-motion resonance, as in some previous discussed cases.

\section{Conclusions}\label{conclu-sec}

In this work we model analytically the orbital angular momentum exchange of two-planet coplanar systems evolving under dissipative forces. Two migration mechanisms were considered: tidal interactions in star-planet and planet-satellite systems and gaseous disc-planet interactions. Our approach is based on the model developed in Michtchenko and Rodr\'iguez (2011), which shows that the angular momentum exchange between the orbital and the exterior components of the total angular momentum can be calculated through the $\alpha$-function, referred to as the orbital angular momentum leakage. For each dissipative mechanism considered, $\alpha$ was calculated as a function of the planet eccentricity of the migrating planet and physical parameters involved in the process.

Using the obtained $\alpha$-function, stationary solutions of the secular conservative problem were calculated for each specific migration scenario and for several values of physical parameters. Stationary solutions provide evolutionary routes that the system would follow in the process of planetary migration and allows us to understand the dynamical history of the system evolving under dissipative forces.

The tidal interactions between a close-in planet with its host star results in the orbital decay of the inner planet. The angular momentum exchange between the orbital component and the stellar rotation imposes constraints on the $\alpha$-function in such a way that $0\le\alpha\le1$. During migration, the planet system always loses some part of the orbital angular momentum, which is used to accelerate the rotation of the star. Large (small) values of the parameter $D$ (ratio between the strengths of planetary and stellar tides) are associated to weak (strong) stellar tides, enabling a substantial conservation (dissipation) of the orbital angular momentum of the system.

In the planet-satellite tidal interaction, the situation is generally opposite. In this case, the migration of the inner satellite is predominantly outward. For typical values of the parameter $D$ (in the range between 1 and 5) and moderate eccentricities ($e_1<0.2$), the $\alpha$-function is always larger than 1. This means that angular momentum is injected into the satellite system, with the planet spin as the supplier source.

Disk-planet interactions modeled through a drag Stokes-like force, have shown that $\alpha\le1$ always. Angular momentum can be removed ($0<\alpha\le1$) or injected ($\alpha<0$) into the planet system. We have shown that the factor $K$ (the ratio between eccentricity damping and orbital decay), which is frequently adopted as a free parameter of the dissipative problem, is a function of the outer planet eccentricity. In addition, $K$ also depends on the parameter $\gamma$, related to the disc properties. Moreover, the assumption of constant $K$ is only valid in the domain of very small orbital eccentricities.

Finally, the development of the $\alpha$-function for each dissipative process, and the consequent calculation of evolutionary routes allow us to
reassemble the starting configurations and migration history of the planet systems on the basis of their current orbital configurations.  In addition, the analysis of the orbital angular momentum evolution during migration of the system allows us to constrain parameters of the involved dissipative physical processes.

\begin{acknowledgements}
This work has been supported by CNPq, FAPESP (2009/16900-5) and INCT-A (Brazil). The authors gratefully acknowledge the support of the Computation Center of the University of S\~ao Paulo (LCCA-USP). We also thank the two anonymous referees for their stimulating revisions.
\end{acknowledgements}

\end{document}